\documentclass{aastex}
\usepackage{emulateapj5}
\usepackage{apjfonts}
\usepackage{graphicx}
\usepackage{epstopdf}

\def\wig#1{\mathrel{\hbox{\hbox to 0pt{%
          \lower.5ex\hbox{$\sim$}\hss}\raise.4ex\hbox{$#1$}}}}

\shorttitle{GJ 436b Transmission Spectra}
\shortauthors{Shabram, et al.}

\newcommand{\me}{$M_{\oplus}$}

\newcommand{\hd}{HD 209458b}

\newcommand{\gj}{GJ 436b}

\newcommand{\cp}{\citep}
\newcommand{\ct}{\citet}

\slugcomment{Revised for ApJ}

\begin{document}

\title{Transmission Spectra of Transiting Planet Atmospheres: Model Validation and Simulations of the Hot Neptune GJ 436b for \emph{JWST}}

\author{Megan Shabram\altaffilmark{1}$^{,}$\altaffilmark{2}, Jonathan J. Fortney\altaffilmark{2}$^{,}$\altaffilmark{3}, Thomas P. Greene\altaffilmark{4}, Richard S. Freedman\altaffilmark{4}$^{,}$\altaffilmark{5}} 

\altaffiltext{1}{Department of Astronomy, University of Florida, 211 Bryant Space Center, Gainesville, FL 32611}
\altaffiltext{2}{Department of Astronomy and Astrophysics, University of California, Santa Cruz, CA 95064}
\altaffiltext{3}{Alfred P. Sloan Research Fellow}
\altaffiltext{4}{Space Science and Astrobiology Division, NASA Ames Research Center, Mail Stop 245-3, Moffett Field, CA 94035}
\altaffiltext{5}{SETI Institute, 515 Whisman Road, Mountain View, CA 94043}

\begin{abstract} 
We explore the transmission spectrum of the Neptune-class exoplanet \gj , including the possibility that its atmospheric opacity is dominated by a variety of nonequilibrium chemical products.  
We also validate our transmission code by demonstrating close agreement with analytic models that use only Rayleigh scattering or water vapor opacity.  We find broad disagreement with radius variations predicted by another published model. For \gj, the relative coolness of the planet's atmosphere, along with its implied high metallicity, may make it dissimilar in character compared to ``hot Jupiters."  Some recent observational and modeling efforts suggest low relative abundances of H$_2$O and CH$_4$ present in \gj 's atmosphere, compared to calculations from equilibrium chemistry.  We include these characteristics in our models and examine the effects of absorption from methane-derived higher order hydrocarbons.  To our knowledge, the effects of these nonequilibrium chemical products on the spectra of close-in giant planets has not previously been investigated.  Significant absorption from HCN and C$_2$H$_2$ are found throughout the infrared, while C$_2$H$_4$ and C$_2$H$_6$ are less easily seen.  We perform detailed simulations of \emph{JWST} observations, including all likely noise sources, and find that we will be able to constrain chemical abundance regimes from this planet's transmission spectrum.  For instance, the width of the features at 1.5, 3.3, and 7 $\mu$m indicates the amount of HCN versus C$_2$H$_2$ present.  The NIRSpec prism mode will be useful due to its large spectral range and the relatively large number of photo-electrons
recorded per spectral resolution element. However, extremely bright host
stars like GJ 436 may be better observed with a higher spectroscopic
resolution mode in order to avoid detector saturation.  We find that observations with the MIRI low resolution spectrograph should also have high
signal-to-noise in the $5 - 10$ $\mu$m range due to the brightness of the
star and the relatively low spectral resolution ($R \sim 100$) of this
mode.
\end{abstract}


\section{Introduction}
The rise of exoplanet characterization since the initial discovery of the transit of planet \hd\ a decade ago \cp{Charb00,Henry00} has been truly stunning.  The initial steps in atmospheric characterization of hot Jupiters came after the realization that the transmission spectra of transiting planets would be diagnostic of the temperature and chemical mixing ratios in these atmospheres \cp{SS00,Brown01,Hubbard01}.  As with perhaps all subfields of astronomy and planetary science, the return on the investment in these myriad observations is amplified when the data sets are compared to models that aim to simulate the conditions in the atmospheres of these planets.

The model atmospheres, whether 1D or 3D, aim to predict the temperature structure, chemical mixing ratios, and wavelength dependent opacity, as a function of height.  A comparison of transmission spectra data with models can enable constraints on the mixing ratios of atomic and molecular absorbers \cp[e.g.][]{Charb02,Fortney03,Tinetti07,Swain08,Sing09b,Madhu09,Desert09,Burrows10}.  Of course it then directly follows that constraints on these atmospheres are \emph{model dependent}, as it is likely that the choices that one makes in constructing a model atmosphere affect the calculated spectrum.  In this exciting field, the great difficulty in obtaining high signal-to-noise observations, along with the somewhat unconstrained nature of current atmosphere models, can make interpretation difficult.  This was again typified by the recent works of \ct{Stevenson10} and \ct{Beaulieu10}, who, based on separate data reductions and model fits, disagree on the probable mixing ratio of methane in the atmosphere of \gj, which we dicuss below.

At the forefront of exoplanet characterization, detections of lower mass objects have expanded the reservoir of constraints to parameter space beyond hot Jupiters, broadening the scope of planetary characterization.  The premier hot Neptune, \gj\, a 22.6 \me\ planet orbiting an M2.5 star, was detected via radial velocity reflex motion by \ct{Butler04}, and later revisited by \ct{Maness07}.  The first photometric detection of transits were obtained by \ct{Gillon07a}, providing the missing link needed to determine the mass and in turn the bulk density.  \ct{Gillon07b} and \ct{Deming07} refined system parameters with a \emph{Spitzer} transit lightcurve at 8 $\mu$m, and \ct{Deming07} and \ct{Demory07} reported a detection of the planet's secondary eclipse at 8 $\mu$m as well.  A recent attempt at transit characterization was \ct{Pont09}, who probed the 1.4 $\mu$m water band using NICMOS on board the \emph{Hubble Space Telescope}.  This was the first  attempt at a multi-wavelength transmission spectrum obtained for \gj.  Ground based efforts have yielded H- and K-band radius measurements \cp{Alonso08,Caceres09}, in addition to radius measurements from EPOXI obtained by \ct{Ballard09} in the 0.35 - 1.0 $\mu$m range.  Using IRAC on-board \emph{Spitzer}, \ct{Beaulieu10} very recently obtained transit depth measurements at 3.6, 4.5 and 8 $\mu$m.

In order to study the dayside of the planet \ct{Stevenson10} obtained secondary eclipse measurements in six \emph{Spitzer} bandpasses, from 3.6 to 24 $\mu$m.  They and \ct{Madhu10} interpret their results as providing evidence of thermochemical disequilibrium in \gj 's dayside atmosphere.  The fits by \ct{Madhu10} to the \ct{Stevenson10} observations are quite interesting in the mixing ratios of CH$_4$, H$_2$O, CO, and CO$_2$ that they require.  They postulate a metal-rich atmosphere, in line with our understanding of Uranus and Neptune.  However, compared to equilibrium chemistry, CH$_4$ is strongly depleted, while H$_2$O is also depleted, and CO and CO$_2$ are strongly enhanced.  Higher order hydrocarbon molecules, e.g., C$_2$H$_{\rm X}$ or HCN, were not considered in their fits, even though these molecules are the first products resulting from methane destruction via photolysis \cp{Moses05} or reduction \cp{Zahnle09b}, and are strong absorbers in the near and mid infrared.  When suggesting that a low CH$_4$ abundance may be due to the photolysis and/or vertical mixing, these higher-order hydrocarbon molecules should be included.  Alternatively, \ct{Beaulieu10} interpret the \ct{Stevenson10} data (some re-reduced by their team), as well as primary transit data, as potentially indicating a methane-rich atmosphere with a temperature inversion.  This only further strengthens the points that differences between data reduction methods and models clearly impacts conclusions.

The coming of the \emph{James Webb Space Telescope} will bring the next advancement in understanding exoplanet atmospheres and exoplanetary atmosphere modeling, as it will help to reduce the under-constrained nature of current models.  For a planet like \gj\, where there are already signs from photometry that the atmospheric chemistry is complex, the spectral capabilities of \emph{JWST} will enhance our understanding, such as the identification of more complex chemical composition regimes and temperature structures.  Understanding the integration times and applicable wavelength bands needed to resolve spectral features of particular atmospheres is key for planning future science initiatives with \emph{JWST}.

\gj\ is a particularly interesting planet because it is the transiting Neptune-class object that orbits the brightest parent star.  This means it may long remain the best studied extrasolar Neptune-class object.  Given the poor signal-to-noise of transmission spectroscopy obtained with \emph{HST} \cp{Pont09}, and the uncertainties that arise from using a small number of wide photometric bands strung together as spectra, it may well fall to \emph{JWST} to enable robust atmospheric characterization.

Deriving accurate transit depths can be a complex task.  Problems with \emph{Spitzer} IRAC primary transit observations include the issue of variable stellar fluxes between transit visits, due to starspots.  In the most common situation where a planet does not occult spots directly, the unocculted parts of the stellar surface appear less bright, due to the spots.  Compared to a stellar surface that lacks spots, one would derive a deeper transit depth \cp[see, e.g.][]{Pont08,Agol10}.  There is also the difficulty of extracting the transit depth with great accuracy, which yields different groups to find different transit depths, with the same data sets \cp[e.g.,][]{Beaulieu08, Desert09}.  However, given that the interpretation of transmission spectroscopy is in principle less sensitive to the atmospheric pressure-temperature profile than day-side emission spectroscopy, transmission spectroscopy may lead to the most robust determinations of atmospheric abundances.

Here we examine the transmission spectrum signatures of a variety of atmospheric chemistries, including those from thermochemical equilibrium calculations, those favored by \ct{Stevenson10}, and those that include abundant higher order hydrocarbons not considered by these authors, guided by results from \ct{Zahnle09b}.  We combine these models with a detailed simulation of the NIRSpec and MIRI Low Resolution Spectrograph instruments that will be aboard \emph{JWST}, for a realistic simulation of what further knowledge we may gain for this important planet.  In \S 2, we make simple comparisons between our models, analytic relations, and other published work, in order to establish the robustness of derived atmospheric parameters, and to validate our code.  We find large differences between our work and that of G. Tinetti and collaborators. In \S 3, we describe model transmission spectra for \gj\, while investigating various chemistries.  In \S 4 we present \emph{JWST} simulations of \gj\ transmission spectra, and discuss the possibilities for identifying important atmospheric signatures in the near future.  In \S 5 we discuss implications of our transmission spectra models as well as address future endeavors for atmospheric characterization via theoretical work.

\section{Transmission Model: Description and Validation}
We model the transmission spectrum of planets using a descendent of the code first described in \ct{Hubbard01}.  Here we ignore the effects of refraction and a glow of photons around the planet's limb due to Rayleigh scattering, both of which \ct{Hubbard01} found to be negligible for close-in planets.  Later works using this code included \ct{Fortney03}, which investigated simple two-dimensional models of the atmosphere of \hd, but included one planet-wide \emph{P-T} profile, with 2D changes in the atmospheric opacities.  \ct{Fortney05c} examined the possible effects of cloud opacity for the slant viewing geometry appropriate for transits.  We refer the reader to \ct{Fortney10}, which described the code in some detail, and extended our treatment to 3D planetary atmosphere models.

In either 1D or 3D, atmospheric \emph{P-T} profiles are lain atop an opaque atmosphere at a reference pressure of either 1 or 10 bar.  The radius at this pressure level is adjusted to yield the best fit to observations.  Along 1000 light ray paths through the atmosphere parallel to the star-planet-observer axis, the local atmospheric density and opacity are each typically sampled at 1000 points along each ray\footnote{Some phrases used in the original description of the code in \ct{Hubbard01} may lead to the impression that the code replaces the true geometry of the atmosphere with a slab having the same column density.  In the \ct{Hubbard01} work this ``slab'' approximation was only done for the simulation of the weak glow of multiply Rayleigh-scattered photons.  This approximation has never been made for 1D or 3D calculations of the transmission spectrum.}  For absorption, the wavelength dependent cross-section is calculated based on contributions from a variety of atoms and molecules.  The abundances can be based either on local chemical equilibrium at a given atmospheric \emph{P--T} point \cp{Lodders02,Lodders06,Lodders09}, or they can be arbitrary.  The Rayleigh scattering cross-section is described in \ct{Fortney10}, and in practice for our H$_2$/He dominated atmospheres, we find a cross-section of 1.645e-24 cm$^2$ molecule$^{-1}$ at 450 nm, and scale by $\lambda ^{-4}$ at other wavelengths.  Here we define the wavelength-dependent transit radius as the radius where the total slant optical depth reaches 0.56, following the results of \ct{Lecavelier08b}\footnote{In our published work he have used two different methods to calculate the ``transit radius.''  In our earliest work, \cp[e.g.][]{Hubbard01,Fortney03} we used the profile of the slant optical depth vs.~radius, set atop an opaque circle, to generate synthetic images from which we calculated the amount of stellar flux blocked by the planet.  We then defined a larger opaque circle whose cross-sectional area blocked this same amount of light as the planetary model---the radius of this larger circle yielded our transit radius. More recently we have chosen the tau=0.56 level as the transit radius, simply for computation ease, but we find fine agreement between the two methods, as did \ct{Lecavelier08b}.  Generally, for simple tests we find excellent agreement between our recent work and that of T.~Barman \ct[e.g.][]{Barman07}, who use a method similar to our current one, as well as with E.~Miller-Ricci Kempton \cp[e.g.][]{MillerRicci09a}.}, who use a method similar to our former one.

\subsection{Opacities} \label{opac}
Opacities generally used in the field of exoplanet atmospheres are discussed in detail in \ct{Sharp07} and \ct{Freedman08}.  \ct{Freedman08} outline the opacities that we use in modeling the atmospheres of hot Jupiters, other Jupiter-class planets, Neptune-class planets, and brown dwarfs \cp[e.g.][]{Fortney08a,Fortney08b,Saumon06,Cushing08}.  We will not repeat the discussions in that paper, but we will touch on the issues of water opacity and higher-order hydrocarbons in turn.

\begin{figure}
\plotone{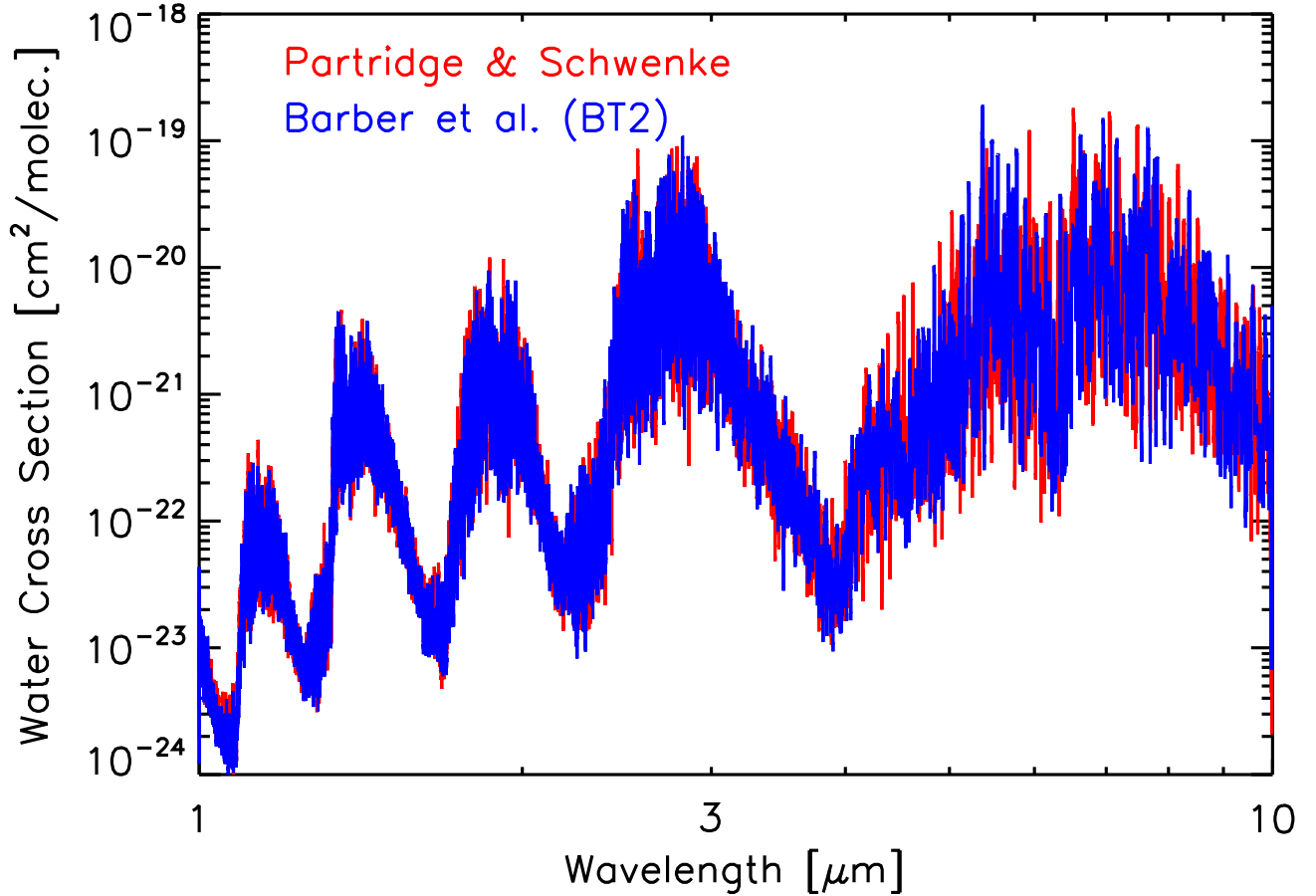}
\caption{Absorption cross-section of water vapor at 1500 K and 1 mbar.  The \ct{Partridge97} opacity is in red while \ct{Barber06} in blue. Differences are very small.}\label{water}
\end{figure}

It is well established theoretically \cp{Burrows97,Marley99,SS00,Sudar00,Barman01} that water vapor opacity is the dominant infrared opacity source in warm giant planet atmospheres.  There is also inescapable observational evidence that this is true for brown dwarfs \cp[e.g.][]{Kirkpatrick05}, and this clearly appears to be true for hot Jupiters as well \cp[e.g.][]{Swain08,Grillmair08}.  Comprehensive \emph{ab initio} calculations of line lists of hundreds of millions of lines for H$_2$O have been tabulated by, for instance, \ct{Partridge97} and by \ct{Barber06}.  Both tabulations are widely used.  One of us (R.~S.~Freedman) has done extensive comparisons of these two particular line lists at the temperatures of interest for planets and brown dwarfs, those below 2500 K, and these differences are described in \ct{Freedman08} at being ``slight.''  In the detailed fits of M.S. Marley and collaborators to L- and T-type brown dwarfs, there is no hint that the \ct{Partridge97} database is insufficient to match the high signal-to-noise medium-resolution NIR and mid-IR spectra that have been achieved for scores of objects \cp[e.g.][]{Cushing08,Stephens09}.  In \mbox{Figure \ref{water}} we show calculated absorption cross-sections at 1500 K and 1 mbar.  Clearly the \ct{Partridge97} and \ct{Barber06} line lists are nearly identical in this pressure/temperature regime.  If transmission spectra calculated by two different atmosphere codes differ (see \S\ref{others}), the choice between these two water line lists cannot be an important contributing factor.

\subsection{Transmission Model Validation} 
\begin{figure}
\epsscale{.75}
\plotone{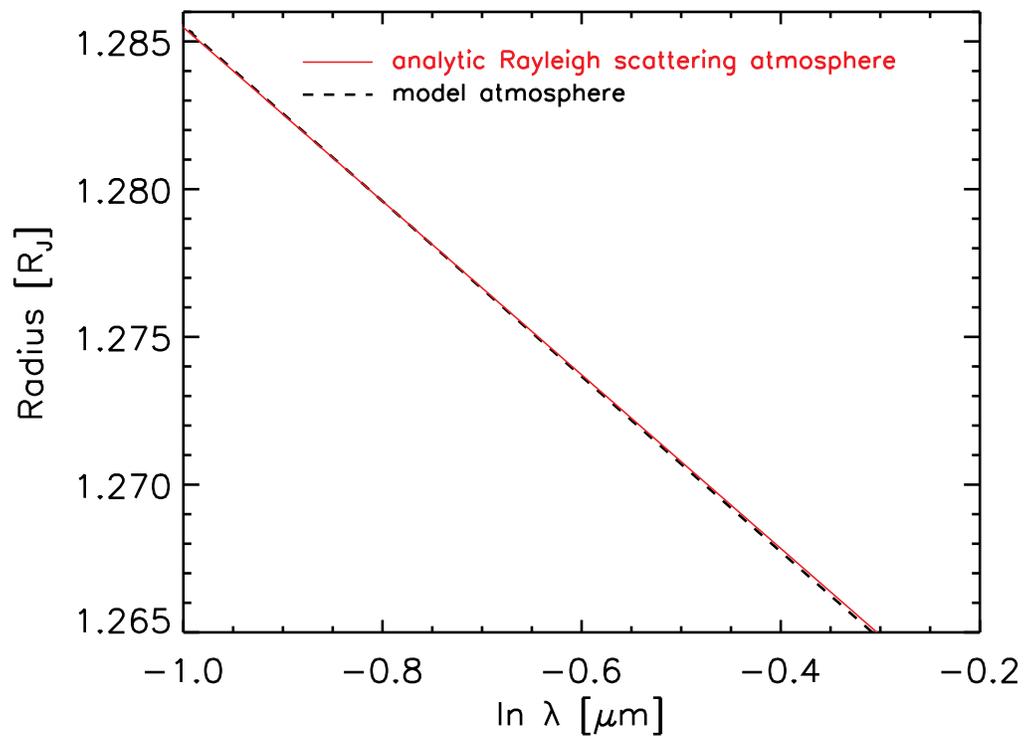}
\caption{Planet radius vs.~wavelength for an isothermal, pure Rayleigh-scattering atmosphere.  The analytical relation of \ct{Lecavelier08a} is shown as the solid line, while our constant-gravity isothermal model is the dashed line.  The slopes of the lines agree to 1\%.
\label{rayl}}
\end{figure}

We can validate the predictions of the transmission spectrum code by turning to previous work.  In particular, \ct{Lecavelier08a} have shown that the relation between absorption cross-section, mixing ratio, atmospheric temperature structure, and transit radius can be treated analytically.  There is a particuarly straighforward relation for the wavelength-dependent transit radius for an atmosphere that obeys a few simple constraints.  These constraints are an isothermal temperature structure, a constant gravitational acceleration with height, and an opacity cross section $\sigma$ that varies as
\begin{equation}
\label{eqalpha}
\sigma = \sigma_o(\lambda / \lambda_o)^{\alpha} ,
\end{equation}
where $\sigma$ and $\sigma_o$ are the wavelength dependent cross-section, and a reference cross-section, respectively, and $\lambda$ and $\lambda_o$ are the wavelength and a reference wavelength, respectively.  Given these constraints, the planet's radius can be written as
\begin{equation}
\label{eqdr}
\frac{dR_{\rm p}}{dln \lambda}=\alpha \frac{kT}{\mu g} = \alpha H,
\end{equation}
where $R_{\rm p}$ is the transit radius, $\lambda$ is the wavelength, $k$ is Boltzmann's constant, $T$ is the temperature, $\mu$ is the mean molecular mass, $g$ is the surface gravity, and $H$ is the scale height.  For a pure Rayleigh scattering atmosphere, $\alpha=-4$.  In Figure \ref{rayl} we present an isothermal model at $T=1500$K, $g=25$ m s$^{-2}$, and $\mu=2.32$, with all opacity turned off, save Rayleigh scattering.  We find a model slope that is within 1\% of the analytic relation from Eq. (\ref{eqdr}), which we regard as excellent agreement. 

\subsection{Comparison With Other Work for Simple Models} \label{others}
\begin{figure}
\centering
  \includegraphics[width=0.75\textwidth]{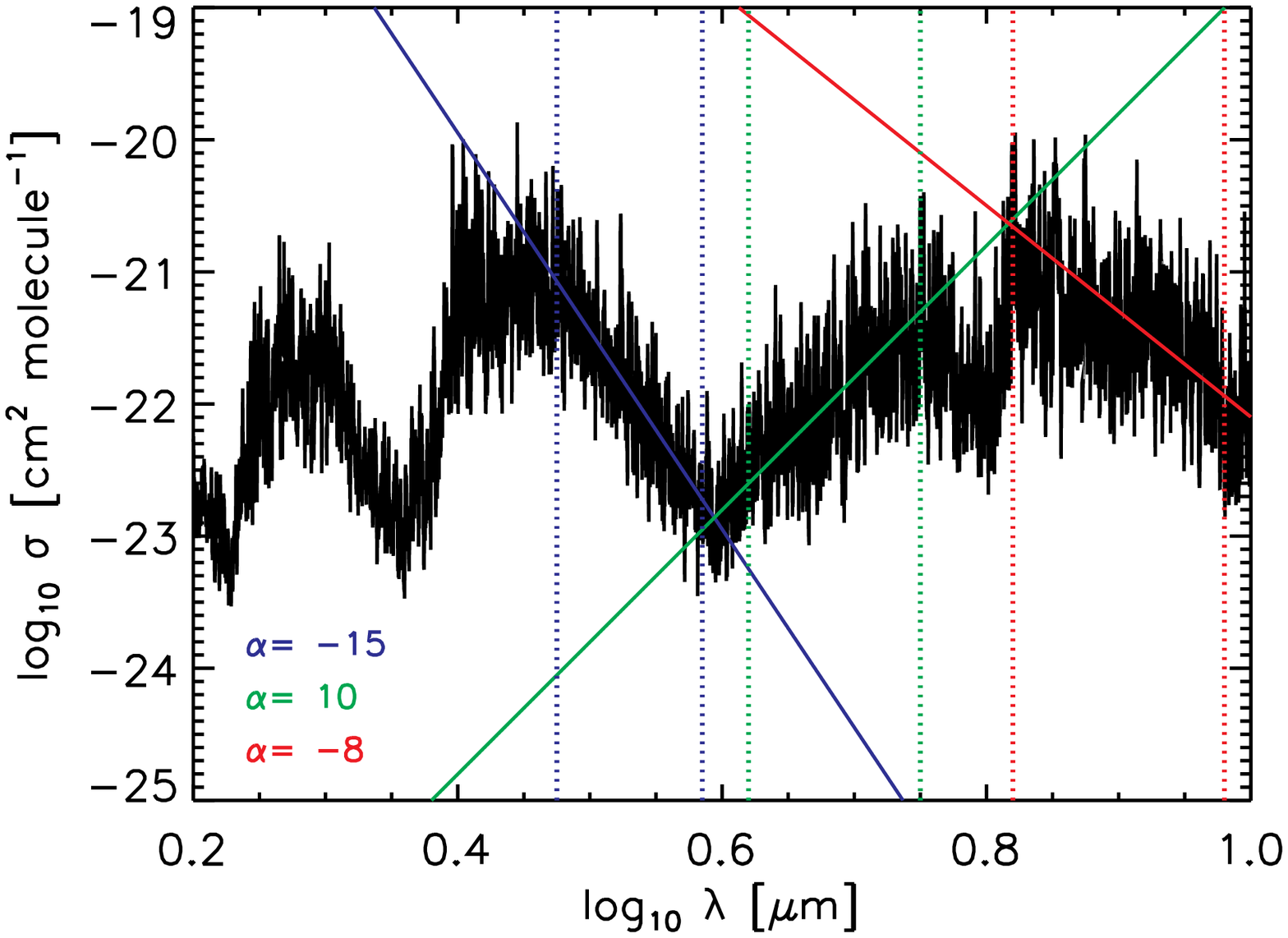}\\
   \includegraphics[width=0.75\textwidth]{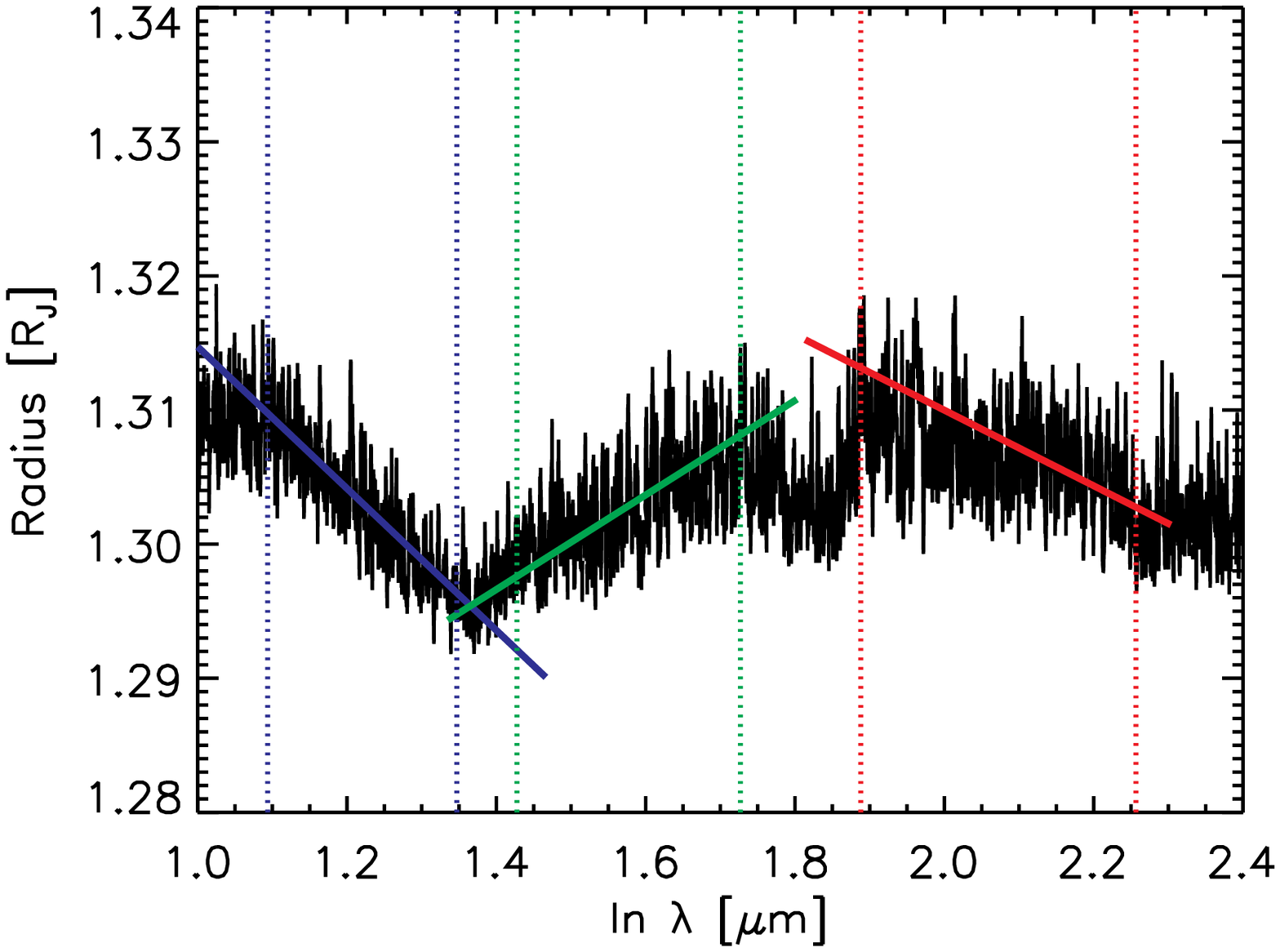}
   \caption{(\emph{a}) Cross-section of water vapor at 1500 K and 1 mbar, with analytic fits to $\alpha$ across three wavelength ranges.  See Equations (\ref{eqalpha}) and (\ref{eqdr}) for details.  Vertical dotted lines show the wavelength range of validity where we expect the analytic relation to hold.  (\emph{b}) The resulting planetary radius vs. ln(wavelength) of the model.  Across the three wavelength ranges, the fit to the linear radius vs. ln(wavelength) is very good.  Again, vertical dotted lines show the range of wavelength validity.\label{waterfit}}
\end{figure}

In \S \ref{opac} we described our implementation of the water vapor line list of \ct{Partridge97}.  As a further test of the transmission code we can isolate specific wavelength regions where water opacity closely obeys the $\alpha$-relation from Eq. (\ref{eqalpha}).  In particular, in Figure \ref{waterfit}\emph{a} we show the absorption cross-section vs.~wavelength at 1500 K and 10 mbar.  We have over-plotted fits for $\alpha$ in three spectral regions.  The bluest and reddest wavelength ranges have a large negative slope, while the middle wavelength range has a positive slope.  If our transmission spectrum model is working correctly, we should be able to match the transit radius slope of $dR_{\rm p}/dln \lambda$ from Equation (\ref{eqdr}), for an atmosphere with a constant gravity and scale height, with water vapor being the only opacity source.  We choose 1500 K and the surface gravity of \hd, 980 cm s$^{-2}$.  This model is plotted in Figure \ref{waterfit}\emph{b}.  One can readily see that over the three defined wavelength ranges that our model matches the analytic relation.

Our choice of an isothermal \hd-like model was based on models presented by G.~Tinetti and collaborators in a recent paper by \ct{Beaulieu09}.  These authors used \emph{Spitzer} IRAC observations to measure the transit depth in 4 bandpasses from 3 to 10 $\mu$m.  Their nice model fit, compared to the data (shown in their Figure 10), allowed the authors to assert that water vapor was the main absorber in that atmosphere.  This may well be true.  However, as shown in \ct{Fortney10}, our own \hd\ models were not able to reproduce the large \emph{variation} in absorption depths.  As discussed in \ct{Fortney10}, we are generally unable to match the much larger variation in transit radius of the models of Tinetti and collaborators \cp[e.g.][]{Tinetti07,Tinetti10}.  Although our two groups use different water opacity databases (Tinetii et al.~use the BT2 list) it does not appear that can be a contributing factor.

\begin{figure}
\epsscale{.75}
\plotone{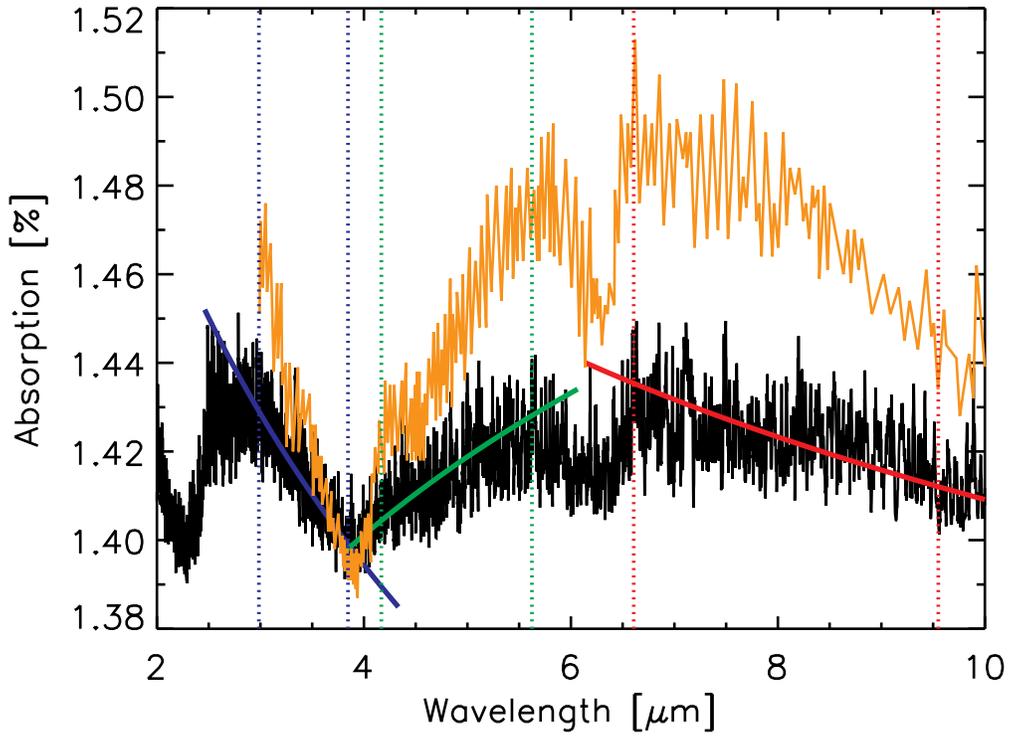}
\caption{Absorption depth vs.~wavelength for a 1500 K isothermal model, with the surface gravity of \hd.  In black is our model from Figure \ref{waterfit}\emph{b}.  The solid-colored curves are the analytic relations from Figure \ref{waterfit}\emph{b} as well.  In orange is a 1500 K, \hd-gravity, model from \ct{Beaulieu09}.  It is readily seen that our model presented here is a substantially better fit to the analytic relation.
\label{compare}}
\end{figure}

To help sort out this issue, we became interested in simple tests.  Figure 9 of \ct{Beaulieu09} additionally shows transmission spectra for isothermal model atmospheres of \hd\ at 1500, 2000, and 2500 K, with opacity due only to water vapor (with a mixing ratio of $4.5 \times 10^{-4}$, the same value we use here).  We compare our 1500 K model to that of \ct{Beaulieu09}, as well as the analytic relations, in Figure \ref{compare}.  (The model from \ct{Beaulieu09} was obtained using a data extraction software package.)  The differences are large.  Here we are able to match the analytic relations, while the model from \ct{Beaulieu09} cannot.

It is not immediately clear what causes these dramatic differences between the two models.  The large differences remain at 2000 K and 2500 K as well.  As shown in Figure \ref{water}, differences in water opacity databases cannot be a culprit.  Differences in abundances can generally not be a reason either, as a higher (lower) water abundance would move the transit radius up (down) at all wavelengths.   Also, that issue was eliminated for the simple test presented here.  We are left in the position of identifying what we believe is a problem, with Tinetti et al. models, but we are not in a position to speculate as to its cause.

We only dwell on this issue at length because it is the match of models to observations that allows for the identification of absorption features, and the determination of the mixing ratios of specific components.  We are left to doubt the validity of the Tinetti transmission models, and the derivations of atmospheric abundances from some of the papers in which those models were used.  We further wish to stress that at this time we have only compared to transmission spectrum models, and not to day-side emission spectrum models.  Certainly the comparison between models is an area in need of future work.  An interesting avenue would be to model the transmission spectrum of solar system planets, such as Earth \cp{Palle09} or Saturn \cp{Nicholson06}, which we will pursue in the near future.  Our description of our methods complete, we can now turn our application to \gj.

\section{Application to GJ 436b}
\gj\ is a relatively small planet with a bulk density similar to Neptune \cp{Gillon07a,Torres08}.  It may be mostly composed of fluid water, but a layer of H-He dominated atmosphere is clearly needed to account for the observed radius \cp{Gillon07a,Adams08,Nettelmann10}.  \gj\ is one of the least-irradiated transiting planets, which makes its atmosphere cooler than many other well-studied planets.  Based on thermochemical equilibrium models, the low temperatures suggests the dominant carbon-bearing molecule in the gaseous envelope is methane \cp{Spiegel10a,Madhu10,Lewis10}.

The recent \ct{Stevenson10} secondary eclipse measurements from warm \emph{Spitzer}, however, have been interpreted by these authors as suggesting otherwise.  An atmosphere who's carbon chemistry is methane-dominate would tend to yield a small flux ratio in the  3.6 $\mu$m band, with more flux in the 4.5 $\mu$m band.  However, \ct{Stevenson10} report a strong detection at 3.6 $\mu$m, and non-detection at 4.5 $\mu$m.  This could be indicative of extensive methane depletion, which would allow one to probe deeply, to hotter gas, in the 3.6 $\mu$m band, while a large mixing ratio for CO and CO$_2$, which both absorb strongly in the 4.5 $\mu$m band, could suppress flux in this bandpass \cp{Stevenson10, Madhu10}.  Large CO and CO$_2$ abundances have been shown to be indicators of high metallicity \cp{Lodders02,Visscher06,Zahnle09a}.

In the favored scenarios, the CO/CH$_4$ and CO$_2$/CH$_4$ mixing ratios are enhanced due to vertical mixing from hotter, CH$_4$-poor gas below, along with the photochemical destruction of CH$_4$ by incident UV photons.  If relatively abundant CH$_4$ is indeed destroyed, this will give rise to a whole host of higher-order hydrocarbons, which is well understood for our solar system's giant planets \cp[e.g.][]{Moses05}.  Detailed chemical models, including photochemistry and vertical mixing, were applied to cool transiting planets by \ct{Zahnle09b}, which predicted the formation of abundant C$_2$H$_2$, C$_2$H$_4$, C$_2$H$_6$, and HCN.  More recently, the theory of a methane-poor atmosphere for \gj\ was challenged by \ct{Beaulieu10}, who, with a combination of modestly different eclipse depths at 3.6 and 4.5 $\mu$m, compute emission and transmission spectra models that allow abundant methane.  

\subsection{Abundances and Opacities at the Terminator} \label{terminator}
\begin{figure}
\epsscale{.75}
\plotone{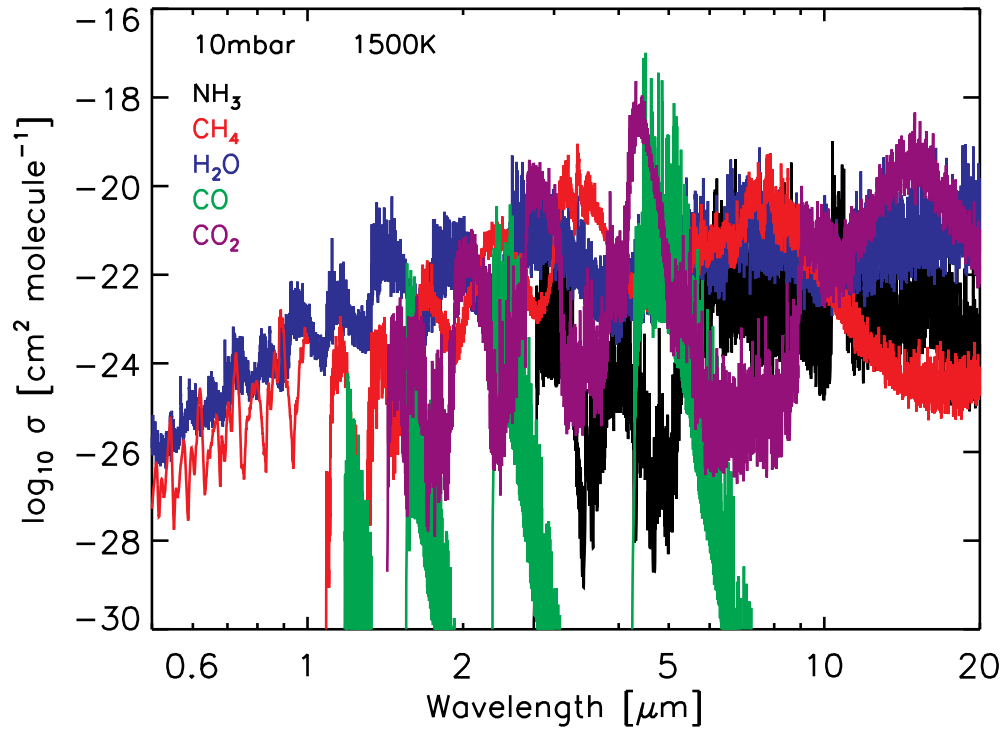}
\caption{Cross sections for the main equilibrium chemistry molecules at 10 mbar and 1500 K.  NH$_3$, CH$_4$, H$_2$O, CO and CO$_2$ are represented by the black, red, blue, green and purple curves respectively.    
\label{eq_x}}
\end{figure}

Since a goal of the work is to explore the prospects for broad-wavelength-coverage spectra with \emph{JWST}, we choose to explore a diverse set of model atmospheres.  Our choices are guided by predictions from equilibrium chemistry, nonequilibrium chemistry, and fits to published spectra of the planet.  Cross sections for the main molecules expected from thermochemical equilibrium are shown in Figure \ref{eq_x}.  There are certainly a rich number of molecular bands, particularly at wavelengths blueward of 5 $\mu$m, where NIRCam and NIRSpec will be sensitive.

\begin{figure}
\epsscale{.75}
\plotone{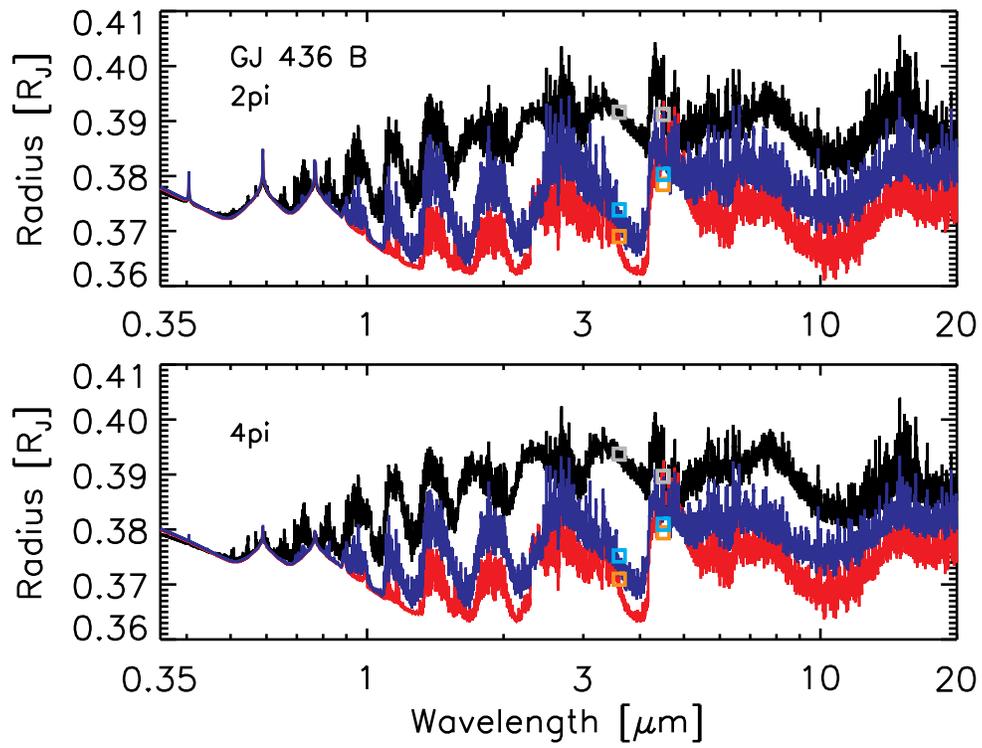}
\caption{Planet radius vs.~wavelength for 2pi (dayside average) and 4pi (planet wide average) \emph{P--T} profiles, at 3 different chemical abundance regimes.  The black models are in chemical equilibrium with 30$\times$ solar metallicity (models `a' and `b' from Table \ref{table1}).  The red and blue models adapt the abundances specified in \ct{Stevenson10} (models `c' through `f'), and are best fits to secondary eclipse measurements at 3.6 and 4.5 $\mu$m.  The Band average radii are plotted for these two bandpasses as squares in grey, cyan and orange corresponding to the black, blue and red models respectively.  
\label{double}}
\end{figure}

\begin{figure}
\epsscale{.75}
\plotone{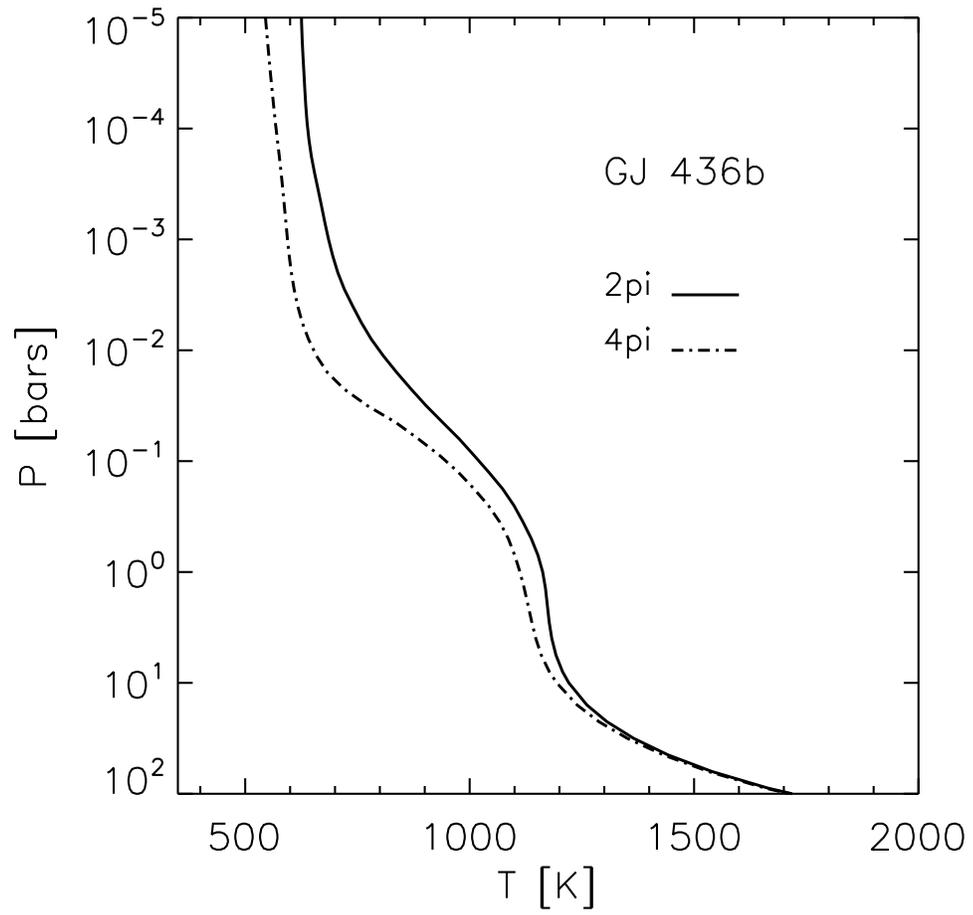}
\caption{Model pressure-temperature profiles for \gj\ used in the current work.  Profiles assume [M/H]=+1.5, or just above 30$\times$ solar.  ``2pi'' is a dayside average profiles while ``4pi'' is a planetwide average. 
\label{pt}}
\end{figure}

In Figure \ref{double}, we present \gj\ model transmission spectra, using these cross sections.  One model is for a 30$\times$ solar metallicity atmosphere in thermochemical equilibrium (black), as well as two models (red and blue) with abundances taken directly from the  ``red model'' and ``blue model'' in Table 2 of \ct{Stevenson10}.  These mixing ratios are somewhat depleted in water, strongly depleted in methane, and CO and CO$_2$-rich compared to our equilibrium calculations.   In all cases, we include the atomic sodium and potassium abundances derived from the equilibrium model.  For the three chemistry cases, we use two different \emph{P--T} profiles.  In the upper panel, a hotter ``dayside average" profile (named ``2 pi'') is used, which simulates inefficient day night energy redistribution.  The bottom panel uses a cooler planet-wide average profile (named ``4 pi'') \cp[e.g.,][]{Fortney05}.  The two \emph{P--T} profiles are plotted in Figure \ref{pt}.  These models (which we will refer to as models `a' through `f'), as well as models discussed in the following sections are described in Table \ref{table1}.  At the current time we do not investigate time-variable temperature structure or abundances.  The only published model of the atmospheric dynamics of \gj\ show little variability \cp{Lewis10}, and there is no observational evidence as yet.  For the two \emph{P--T} profile cases, we adjust the 10 bar radius of all models so that the radii align in the optical.  The main chemical difference between the hotter and cooler profiles is that the hotter model yields a larger mixing ratio of Na and K, yielding stronger features in the optical.  A structural difference is that the larger scale height in the warmer model leads to modestly larger changes in radius as a function of wavelength.

Compared to our equilibrium model, the two models from \ct{Stevenson10} (models `c' through `f' in Table \ref{table1}) show considerably smaller radii in the near and mid-infrared, which is predominantly due to the smaller mixing ratio of water.  Therefore, a clear probe of the water mixing ratio is the near-IR radius, compared to the optical radius, where water vapor is much less important.  As expected the differences between the equilibrium model and the red/blue models are greatest in wavelengths where methane and water are the dominant absorbers, and smallest where CO and CO$_2$ are the dominant absorbers.

\subsection{Nonequilibrium Chemical Products:  Absorption Features} \label{photochem}
\ct{Stevenson10} and \ct{Madhu10} suggested that there could be evidence for a very low CH$_4$ mixing ratio in the atmosphere of \gj, due to a combination of vertical mixing and photolysis of CH$_4$.  However, they did not investigate how the methane-derived nonequilibrium chemical products may affect the spectra of the planet's atmosphere.  Recently, \ct{Zahnle09b} have investigated nonequilibrium carbon for isothermal ``warm Jupiter'' atmospheres ($800<T<1200$ K) at a range of metallicities.  These temperatures are similar to those suggested for \gj.  \ct{Zahnle09b} find that methane is sustained at higher regions in the atmosphere, water is more stable, and OH and H$_2$ quickly combine to form H$_2$O and H.  This effectively increases the C to O ratio leading to increased abundances of molecules such as HCN, C$_2$H$_2$, C$_2$H$_4$, and C$_2$H$_6$.  Given the uncertainties in modeling this chemistry, and the wide range of atmosphere models that are consistent with the \gj\ data to date, our next aim is to explore transmission spectra with a range of nonequilibrium chemical products, guided by the results of \ct{Zahnle09b}.

\begin{figure}
\epsscale{.75}
\plotone{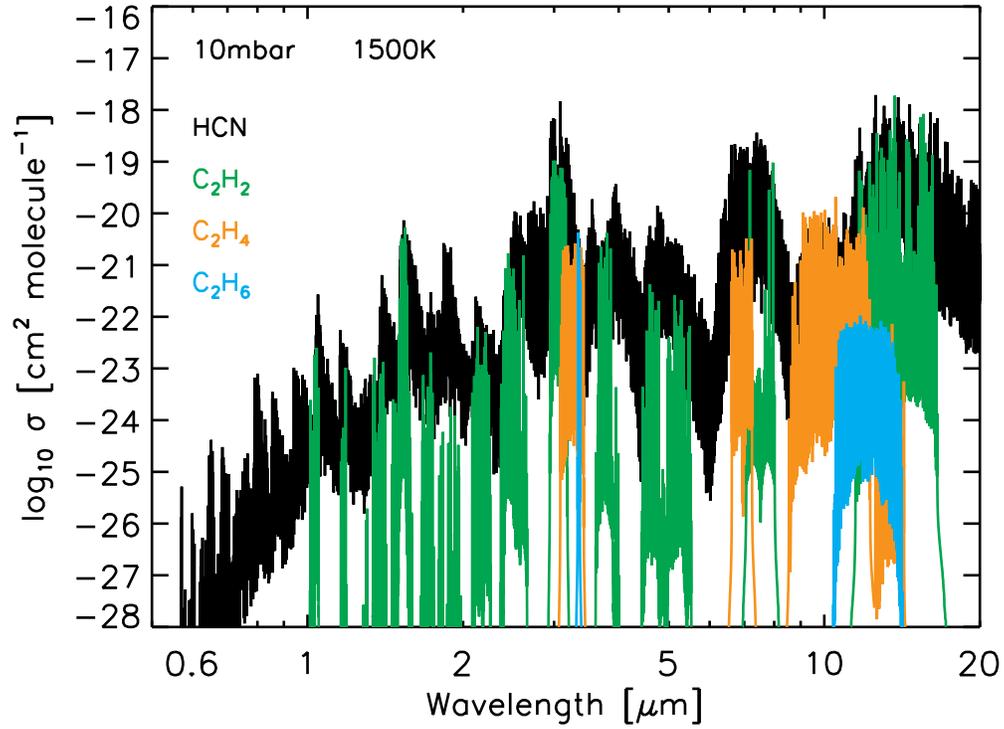}
\caption{Cross sections for the main higher-order hydrocarbon products at 10 mbar and 1500 K.  HCN, C$_2$H$_2$, C$_2$H$_4$ and C$_2$H$_6$ are represented by the black, green, orange and cyan curves respectively.    
\label{photo_x}}
\end{figure}

\begin{figure}
\epsscale{.75}
\plotone{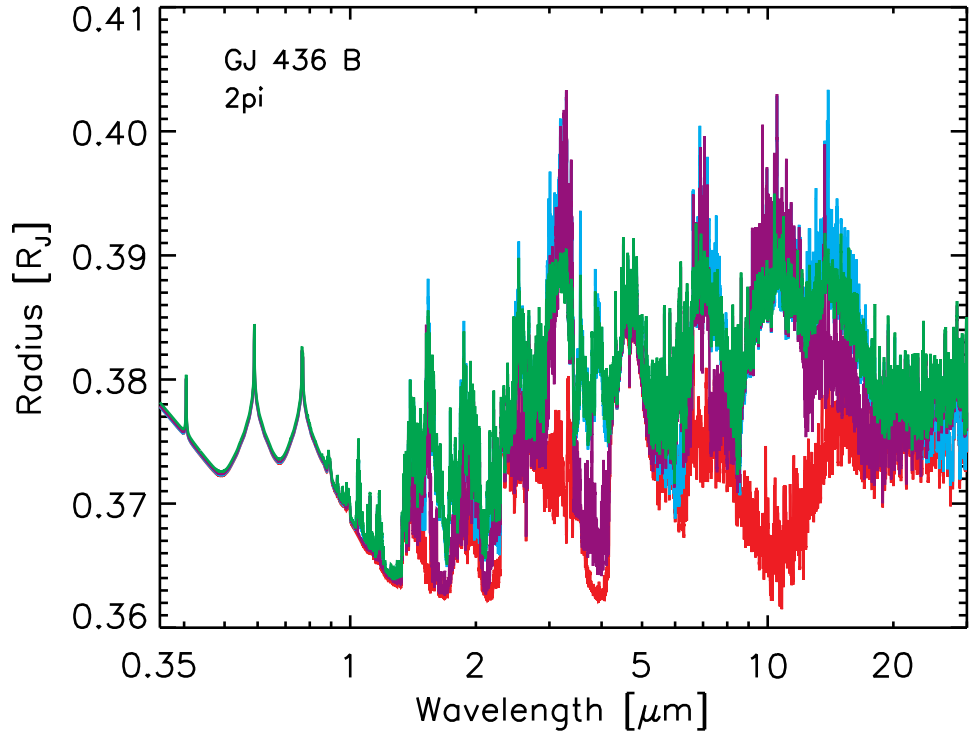}
\caption{Planet radius vs.~wavelength using a 2pi (dayside average) \emph{P-T} profile for the `red' model described in \ct{Stevenson10} shown in red (model `e' from Table \ref{table1}.  Also shown are `red' models with additional absorption due to nonequilibrium chemical products.  In purple is model `e' including absorption from HCN, C$_2$H$_2$, C$_2$H$_4$, and C$_2$H$_6$ with abundances of $1 \times 10^{-4}$, $1 \times 10^{-5}$, $1 \times 10^{-3}$ and $1 \times 10^{- 8}$ respectively (model `g' from Table \ref{table1}).  In cyan depicts this model again, removing absorption from HCN, revealing C$_2$H$_2$ as the dominant feature at 3.3 $\mu$m (model `h' from Table \ref{table1}).  In green we include chemical abundances that become absent above 10 mbar, a condition adapted from \ct{Zahnle09b} (model `i' from Table \ref{table1}).  
\label{photo}}
\end{figure}

The cross sections for these first generation products of methane, which are strong absorbers in the infrared, can be found in Figure \ref{photo_x}.  Data for C$_2$H$_2$, C$_2$H$_4$ and C$_2$H$_6$, which are likely incomplete, are from the HITRAN database, while that for HCN is from the calculations of \ct{Harris08}.  The cross-sections are similar in magnitude to those of the equilibrium chemistry products (Figure \ref{eq_x}).  HCN and C$_2$H$_2$ have prominent features at 1.5, 3.3, 7, and 13 $\mu$m, C$_2$H$_4$ has a significant impact to absorption at 9.5 $\mu$m, but C$_2$H$_6$ has very little effect, with most of its opacity residing between 10 and 13 $\mu$m.  In Figure \ref{photo} we explore transmission spectra including the opacities with these nonequilibrium products. We begin with model `e' in from Table \ref{table1}, also shown as the red model from the top panel of Figure \ref{double}.  This model uses a dayside average \emph{P--T} profile with the same best fit mixing ratios as the ``red model'' from Table 2 of \ct{Stevenson10}.

In addition to the original parameters of model `e', we include absorption from HCN, C$_2$H$_2$, C$_2$H$_4$ and C$_2$H$_6$, with mixing ratios of $1 \times 10^{-4}$, $1 \times 10^{-5}$, $1 \times 10^{-3}$ and $1 \times 10^{- 8}$ respectively, shown as the cyan model in Figure \ref{photo} (model `g' from Table \ref{table1}).  We derive these chemical abundances based on \ct{Zahnle09b}, where they depict mixing ratios as a function of height and $K_{\rm zz}$ (eddy diffusion coefficient) for an atmosphere at 1000 K.  The mixing ratios chosen are similar to what is expected for an atmosphere that is vigorously mixing ($K_{\rm zz}=10^{11}$).

It is worthwhile to explore some variations on this model.  One variation removes the blanketing effects of hydrogen cyanide (HCN), shown in purple (model `h' from Table \ref{table1}).  This model illustrates the difference in the width of the features at 3.3, 7 and 13 $\mu$m, where HCN and C$_2$H$_2$ have overlapping opacity.  When the features are thick, they are dominated by HCN opacity.  When we remove HCN, we see thinner features, indicating that C$_2$H$_2$ dominates over HCN.  This characteristic may allow further constraints on the mixing ratios of cool transiting planet atmospheres. In green is a model with these nonequilibrium products absent at pressures below 10 mbar (model `i' from Table \ref{table1}).  The green model illustrates the condition of modest vertical mixing, with $K_{\rm zz}$ on the order of $10^{6}$ cm$^2$ s$^{-1}$, similar to that favored by \ct{Madhu10}.  In this case, nonequilibrium products are not stable higher in the atmosphere than the $\sim$~10 mbar pressure level \cp{Zahnle09b}.  The main absorption features in the infrared will be weakened, as the green model confirms.  The strength of these features could be constraints on vertical mixing and $K_{\rm zz}$.

Absorption from these nonequilibrium products may mask features that would otherwise show an under-abundance of water in the transmission spectrum, as Figure \ref{photo} shows.  The red model, which generally shows smaller radii in the near- and mid-IR than in the optical, instead shows significantly larger radii when the nonequilibrium products are introduced.  Spectroscopy, rather than photometry, will be key towards disentangling the effects of various molecules on the transmission spectrum.

\subsection{Comparison with Data} \label{comparetext}
\begin{figure}
\epsscale{.75}
\plotone{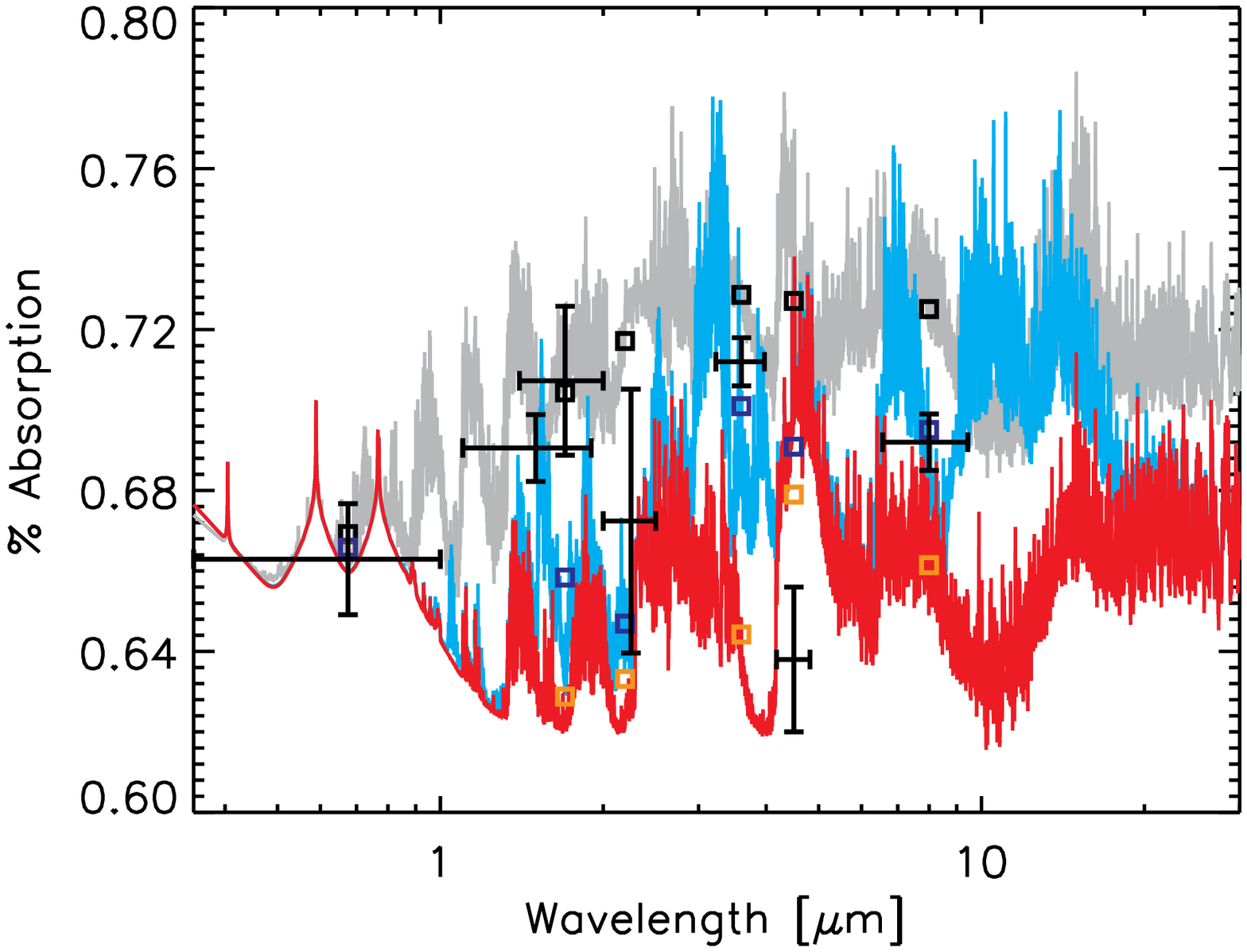}
\caption{We present model fits to data from  \ct{Ballard09} (0.35 - 1.0 $\mu$m from EPOXI), \ct{Pont09} (1.1 - 1.9 $\mu$m using \emph{HST NICMOS}), \ct{Alonso08} (ground based H-band), \ct{Caceres09} (ground based K-band) and \ct{Beaulieu10} (\emph{Spitzer} IRAC 3.6, 4.5, and 8 $\mu$m bands).  The grey model is a 30$\times$ solar metallicity model with a 2pi \emph{P--T} profile (model `a' from Table \ref{table1}).  We show the red and cyan models from Figure \ref{photo} again here (models `e' and `g' from Table \ref{table1}, respectively).   It is clear that we are unable to reproduce the large peak to trough variation in spectra that allowed \ct{Beaulieu10} to assert reasonable agreement of their models to current data.  
\label{data}}
\end{figure}

In Figure \ref{data}, we explore model fits to recent \emph{Spitzer} IRAC data for 3.6, 4.5, and 8 $\mu$m \cp{Beaulieu10}, as well as data from EPOXI in the 0.35 - 1.0 $\mu$m range \cp{Ballard09}, \emph{HST NICMOS} in the 1.1 - 1.9 $\mu$m range \cp{Pont09}, ground based H-band \cp{Alonso08}, and ground based K-band \cp{Caceres09}.  We plot our models in high resolution, and as band-averages, where appropriate.  Looking at the optical and near infrared, the best match is to the model that uses abundances from equilibrium chemistry, shown in grey (model `a' from Table \ref{table1}).

In the mid-infrared, for the \emph{Spitzer} IRAC data we see the same trend that we found in \ct{Fortney10}:  in comparison to \ct{Beaulieu08} data for HD 189733b and to \ct{Beaulieu09} data for \hd, our models cannot match the larger \emph{amplitude} of the features--the implied dramatic change in absorption depth as a function of wavelength.  The models of G.~Tinetti and collaborators, used in the \ct{Beaulieu10} paper, do fit the observation reasonably well.  However, as discussed in \S 2.3, we find that the Tinetti et al.~models overestimate the amplitude of absorption features.

It is clear that our models do not agree with the large peak to trough variation of spectra required by \ct{Beaulieu10} to fit their data.  If the methods employed by \ct{Beaulieu08,Beaulieu09,Beaulieu10} are correct, and the error bars are not underestimated, the results for all of these planets imply that dramatic revisions to models of these atmospheres are needed.  Given the uncertainties in the reduction of IRAC transit data, it may be up to \emph{JWST} to confirm speculations about the molecules present at the terminator of \gj.  

\section{Prospects for JWST for GJ 436}
We now evaluate the observability of the differences in the \gj\ 
models by comparing them through the eyes of \emph{JWST}. We have developed a code that simulates \emph{JWST}
spectra by computing the number of photons detected using a model of the
host star, a transmission model of the planet, and estimates of the total
efficiency (detected electrons per incident photon) at each wavelength for the
various \emph{JWST} dispersive spectroscopic modes. Noise is also modeled and
added to the simulated spectra.

The star GJ 436 is relatively bright over the $0.7 - 5$ $\mu$m spectral
region, and the transmission models predict that \gj\ will have absorption features
from many species over this wavelength range. Therefore we illustrate the
model similarities and differences with simulations of \emph{JWST} observations
using the NIRSpec $R \equiv \lambda / \delta \lambda \sim 100$ spectroscopic
mode over this spectral range. The double-pass CaF$_{2}$ prism used in this
mode provides spectroscopic resolution varying from $R \simeq 30$ at
$\lambda \sim 1.2$ $\mu$m to $R > 200$ at $\lambda > 4.3$ $\mu$m.  We
approximate this with a fourth order polynomial fit over the $0.7 - 5$
$\mu$m range, and we assume that the prism has total transmission efficiency of
0.81 after two passes. We estimate that the optical efficiency of NIRSpec's 14
reflective surfaces \citep{tePlate05} is approximately 0.58 over $\lambda =
1 - 5$ $\mu$m, consistent with the values calculated by the NIRSPec team (P.
Jakobsen, private communication 2003) and assumed by \citet{Deming09}
after removing the grating blaze function.  We adopt a quantum efficiency of
0.75 across the entire $\lambda = 1 - 5$ $\mu$m spectral range, consistent
with the NIRSpec detector requirements \citep{BRauscher07}. The telescope is estimated to have total reflectivity of 0.9 across this wavelength range.  Total
efficiency was modeled to decrease linearly by a factor of 2.0 as wavelength
decreases from 1.0 to 0.7 $\mu$m, driven mostly by reduction in reflectivity
in the 14 reflective NIRSpec surfaces.

Like \citet{Deming09}, we assume there will be no losses from the 1\farcs6 wide entrance slit and that the only significant noise sources are photon noise and systematic noise due to
small guiding errors during exposures. Photon noise is simulated by adding
Poisson noise appropriate for the number of detected photo-electrons in each
resolution bin. We adopt the systematic noise value of $5 \times 10^{-5}$
estimated by \citet{Deming09}. We do assume that this noise is Gaussian
in its distribution although \citet{Deming09} found that it was somewhat
non-Gaussian. Even with high precision JWST instruments, we will suffer systematic noise at these modest but significant levels. In the "1-(in-transit/star)" computation, any additional natural or instrumental noise occurring at frequencies greater than the inverse of the transit observation period will impact the extracted spectrum.  This simulation program was coded in C, and it uses the public
domain RANLIB package for simulating photon noise and Gaussian systematic
noise.

A high fidelity stellar model of the GJ 436 host star was not readily
available, so we used a model of GJ 411 which has M2 V spectral type,
similar to the M2.5 V type of GJ 436. Using our simulation code, we
re-binned the \citet{Kurucz09} $R = 1000$ model of GJ 411 to the
instrumental resolution of the \emph{JWST} NIRSpec prism at each wavelength
interval over $\lambda = 0.7 - 5$ $\mu$m. Next, our code computed the number
of stellar photons from this binned flux, reducing it by the ratio of the
squared model planet radius divided by the squared stellar radius (assumed
to be $3.2 \times 10^{10}$ cm) at each wavelength. We used a distance of
10.2 pc to GJ 436 and an integration time of 1800~s for these calculations.
This integration time is $\sim 33\%$ shorter than the 2740 s duration of the
transit \citep{Pont09}. We used the resultant simulated in-transit spectrum and the
simulated stellar spectrum of equal integration time to compute the
absorption depth at each wavelength, 1 - (in-transit / star). This is
plotted for \gj\ models in Figure \ref{jwst}.

Models `a', `c', and `e' (shown in black, blue, and red respectively) are plotted in the top panel of Figure \ref{jwst}.  These are \emph{JWST} simulations of the models in the top panel of Figure \ref{double}.  Models `e', `g', `h', and `i' are plotted in the bottom panel in red, purple, cyan and green respectively, and are \emph{JWST} simulations of the models in Figure \ref{photo}.  The absorption features are labeled here for clarity.  In particular, in the top panel the differences between the water rich model (black) and water poor models (red and blue) are readily apparent, as is the strong CO$_2$ feature at 4.3 $\mu$m and CO feature at 4.5 $\mu$m.  In the bottom panel, the absorption features due to nonequilibrium HCN and C$_2$H$_2$ are clearly apparent, as is CH$_4$ absorption from 3-4 $\mu$m.  The prospects for detailed characterization of this planet, and others with \emph{JWST}, is good.

Given the high brightness
of \gj\ ($K = 6.1$ or $K_{\rm AB} = 7.9$ mag), it is likely that its
observation will require use of a detector subarray that is smaller in the
dispersion direction than the $\sim 350$ pixel length of the complete $0.6 -
5$ $\mu$m $R \sim 100$ spectrum \citep{Tum08}. Therefore, we find that acquiring the entire spectrum shown at the signal-to-noise in Figure \ref{jwst} may require 2 or 3 transits.  These
observations may be best acquired in the higher resolution $R = 1000$
mode for stars as bright as GJ 436. 

At $\lambda = 5 - 10$ $\mu$m MIRI low resolution spectrograph (LRS; R $\sim$ 100) observations were also simulated for these models in a similar fashion using details of LRS models and actual measured performance. These simulations are shown in Figure \ref{ir} and are another independent way of looking at the effects of nonequilibrium chemistry in the atmosphere of \gj.  In the top panel (models `a', `c' and `e' are shown as black, blue, and red respectively), water is the main opacity source in this wavelength range.  The bottom panel (showing models `e', `g', `h', and `i' as red, purple, cyan, and green respectively) shows the clear distinction of HCN and C$_2$H$_2$ between the different models shown at 7 $\mu$m.  Absorption from C$_2$H$_4$ is shown at 9.5 $\mu$m as well.  Observations using MIRI will be the only way to probe the 9.5 $\mu$m feature created by the presence of C$_2$H$_4$.  The simulations were also made for a total integration
time of 30 m in transit and 30 m on the star. The flux of GJ 436 is
less than 1 Jy over this wavelength range, faint enough for its entire
$\lambda = 5 - 10$ $\mu$m spectrum to be acquired simultaneously, which is an advantage MIRI will have over NIRSpec observations that require multiple transits to obtain the full spectrum.    

The noise included in the spectral simulations and shown in Figures \ref{jwst} 
and \ref{ir} are likely lower limits to the actual noise recorded in \emph{JWST}
spectra. Common-mode low frequency noises due to the observatory will
likely be eliminated by the differential measurement of the star and
planet. All data for the simulated spectra shown are obtained on time
scales of 2 -- 3 eclipses, so any global variation in the star on
longer time scales (i.e., subsequent transits or eclipses) will also be
removed. Variation of the planet between transits will be
recorded, and this could, in principle, limit the usefulness of co-adding multiple
spectra. Variations in the \emph{JWST} instrumentation, the host star, or the
planet on time scales shorter than 2 -- 3 transit or eclipse events will
appear in the data as systematic or random noise. Nevertheless, we
expect that \emph{JWST} should be able to obtain high signal-to-noise
exoplanet spectra given that \emph{Spitzer} was able to obtain signal-to-noise
approaching $10^4$ \citep[e.g.,][and references therein]{Machalek10}
with its older detectors, about a factor of 2 less than we predict for
the systematic noise of \emph{JWST}.

\section{Discussion and Conclusions}
We have explored possible transmission spectra of \gj.  We have investigated equilibrium chemistry cases and have compared them to mixing ratios derived to fit recent secondary eclipse measurements in the near infrared.  The differences in these mixing ratio regimes have a significant effect when comparing the corresponding transmission spectra.  The lack of water in some cases will be easily detected by \emph{JWST} assuming the spectrum is not dominated by absorption from nonequilibrium chemical products of methane.  For this reason, we have made the first attempt to include opacity from higher order hydrocarbons for close-in giant planet atmosphere models.  Specifically, we have included absorption from HCN, C$_2$H$_2$, C$_2$H$_4$, and C$_2$H$_6$.  The special condition that favors production of these molecules the most is the relatively low temperatures found in this planet.

\emph{HST} and \emph{Spitzer} may be limited in the capabilities needed to observe this planet's transmission spectrum.  The \ct{Pont09} near-IR spectrum from \emph{Hubble} suffered from instrumental systematics that could not be overcome.  Reduction of \emph{Spitzer} IRAC transit data suffers from the problem that different groups reducing the same data sets achieve different results, with errors bars that do not overlap \cp{Beaulieu08,Desert09}.  Transit depths observed at different epochs allow for stellar vairability to complicate the interpretation.  These observations are \emph{extremely challenging}, such that many results for \gj, and many of the transiting planets, must be regarded as provisional.

\begin{figure}
\epsscale{.75}
\plotone{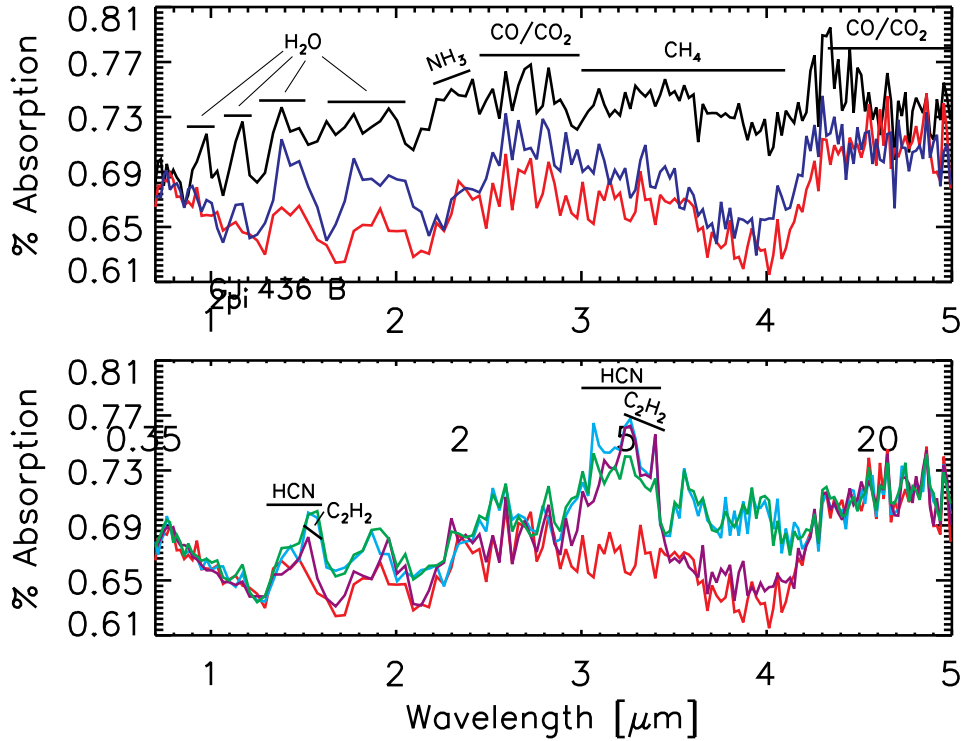}
\caption{Simulated \emph{JWST} NIRSpec prism mode observations of the absorption
depth of the transiting planet GJ 436b relative to its host star at each
wavelength. An integration time of 1800~s was used for both the in-transit
and star-only spectra simulations.  The models shown were previously plotted in Figures \ref{double} and \ref{photo}.  Models `a', `c', and `e' are shown in black, blue, and red respectively.  Models `e', `g', `h', and `i' are shown in red, purple, cyan and green respectively.  It is clear that we will be able to understand complex mixing ratio regimes with \emph{JWST}.  
\label{jwst}}
\end{figure}

\begin{figure}
\epsscale{.75}
\plotone{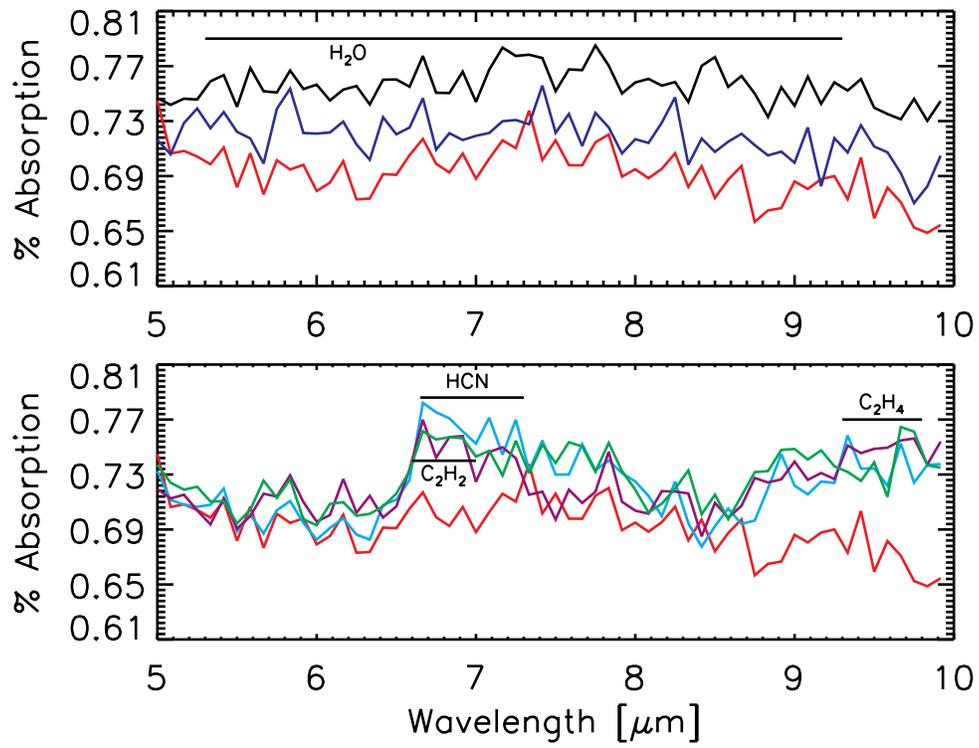}
\caption{Simulated \emph{JWST} MIRI Low Resolution Spectrograph observations of the absorption
depth of the transiting planet GJ 436b relative to its host star at each
wavelength.  Models `a', `c', and `e' are shown in black, blue, and red respectively.  Models `e', `g', `h', and `i' are shownl in red, purple, cyan and green respectively.In the top panel, water is the main absorber in this wavelength range.  In the bottom panel, HCN and C$_2$H$_2$ features are labeled at 7 $\mu$m, and C$_2$H$_4$ absorption is labeled at 9.5 $\mu$m for clarity.  
\label{ir}}
\end{figure}

We have shown that \emph{JWST} should be able to make important breakthroughs in answering questions about the chemical mixing ratios in the planet's atmosphere. Specifically, the presence of higher order hydrocarbons in the transmission spectrum of this planet would reinforce the recent claim of non-detection of methane in the emission spectrum.  We also may be able to constrain the composition and structure at the planetary limb.  If hydrocarbon chemistry is present, it will be possible to constrain the abundances of these species.  For example, HCN and C$_2$H$_2$ have strong absorption overlapping at 1.5, 3.3, 7, and 13 $\mu$m. The thickness of the feature in each wavelength region will tell us which molecule is more abundant in the atmosphere.  When there is minimal HCN, the signature of C$_2$H$_2$ in these wavelength regions will be characteristically thinner. This phenomena is depicted for \emph{JWST}'s NIRSpec in the lower panel of Figure \ref{jwst} for the 1.5 and 3.3 $\mu$m features as well as for the MIRI LRS in the lower panel of Figure \ref{ir} at 7 $\mu$m.  The strength of the features in the infrared produced by these hydrocarbons may also be instrumental in constraining information regarding atmospheric mixing.  Weak features may indicate a small eddy diffusion coefficient ($K_{\rm zz}$), pointing towards weak vertical mixing in the planet's atmosphere, with these molecules forming at higher pressure levels.

In order to produce accurate model spectra of an exoplanet, including nonequilibrium chemical products (due to mixing and photochemistry) a self consistent code is needed.  This would ensure that the molecular mixing ratios, and their impact on opacities, are consistent with the atmospheric \emph{P--T} profile.  This alone is difficult, but should be addressed in the future.  In principle, all of this should be done in a three-dimensional model.  Perhaps someday exquisite observations will warrant such a  treatment.  Clearly \emph{JWST} will be an effective tool for reducing the under-constrained nature of current models, yet the brightness of GJ 436 may be a limitation.  The NIRSpec prism mode will be able to observe a larger spectral range at high signal-to-noise with fewer transits for planets orbiting stars that are relatively dimmer than GJ 436.  This issue however, does not pertain to obtaining spectra in the 5 - 10 $\mu$m range using the MIRI LRS.  This study is the first in a series of papers that will investigate models of transmission and emission spectra of transiting planets, convolved with our realistic model of \emph{JWST} observations.  For a complementary approach over a large phase space of \emph{JWST} exoplanet observations, see \ct{Belu10}.

The model dependent nature of our understanding of the composition and structure of exoplanet atmospheres ensures that it will always be useful to refine atmosphere models.  Thus developing, scrutinizing and refining these models brings us closer to the true nature of these planets.  In \S2 we validated our model against analytic relations, but found that a similar Tinetti et al. model could not match these same relations. Recent work on including nonequilibrium and photochemical products \cp{Liang04,Zahnle09a,Zahnle09b} will yield better predictions of chemical mixing ratios.  Transmission spectrum models that incorporate the full 3D nature of the planetary atmosphere, which is particularly imporant at the terminator region \cp{Fortney10,Burrows10} will yield a more accurate understanding of the temperature structure and chemical abundances.  The conditions of \gj\ coupled with \emph{JWST} will provide the chance to explore the possibility of a compositionally complicated atmosphere in great detail, expanding our knowledge of planets as unique to their particular circumstance.

\acknowledgements
We thank Jean-Michel D{\'e}sert, David Sing, David Spiegel, Alain Lecavelier des Etangs, Mark Marley, Heather Knutson, and Adam Burrows for useful discussions, and Eliza Kempton, Travis Barman, and Nikku Madhusudhan for discussions and the sharing of model results.  J.~J.~F.~ and M.~S.~acknowledge the support of the \emph{Spitzer} Theory Program and a University Affiliated Research Center (UARC) Aligned Research Program (ARP) grant.  UARC is a partnership between University of California, Santa Cruz and NASA Ames Research Center.  T.~P.~G.~acknowledges support from the JWST NIRCam instrument, NASA WBS 411672.05.05.02.02.


\pagebreak
\begin{table}[ht]
\caption{Model Transmission Spectra for GJ 436b}
\centering
\begin{tabular}{ c c c c c c c c c c }
 \\ [0.5ex]
\hline\hline 
 & & \multicolumn {8} {c} {Abundances} \\
\cline {3-10}
Model & Profile & H$_2$O & CH$_4$ & CO & CO$_2$ & HCN & C$_2$H$_2$ & C$_2$H$_4$ & C$_2$H$_6$  \\ [0.5ex]
\hline
a & 2pi &  \multicolumn {8} {c} {$ 30 \times Solar $} \\
b & 4pi &  \multicolumn {8} {c} {$ 30 \times Solar $} \\
c & 2pi & $1 \times 10^{-4}$ & $1 \times 10^{-7}$ & $1 \times 10^{-4}$ & $1 \times 10^{-6}$ & --- & --- & --- & ---  \\
d & 4pi & $1 \times 10^{-4}$ & $1 \times 10^{-7}$ & $1 \times 10^{-4}$ & $1 \times 10^{-6}$ & --- & --- & --- & --- \\
e & 2pi & $3 \times 10^{-6}$ & $1 \times 10^{-7}$ & $7 \times 10^{-4}$ & $1 \times 10^{-7}$ & --- & --- & --- & ---  \\
f  & 4pi & $3 \times 10^{-6}$ & $1 \times 10^{-7}$ & $7 \times 10^{-4}$ & $1 \times 10^{-7}$ & --- & --- & --- & ---   \\
g & 2pi & $3 \times 10^{-6}$ & $1 \times 10^{-7}$ & $7 \times 10^{-4}$ & $1 \times 10^{-7}$ & $1 \times 10^{-4}$ & $1 \times 10^{-5}$ & $1 \times 10^{-3}$ & $1 \times 10^{-8}$   \\
h & 2pi & $3 \times 10^{-6}$ & $1 \times 10^{-7}$ & $7 \times 10^{-4}$ & $1 \times 10^{-7}$ & --- & $1 \times 10^{-5}$ & $1 \times 10^{-3}$ & $1 \times 10^{-8}$   \\
i & 2pi & $3 \times 10^{-6}$ & $1 \times 10^{-7}$ & $7 \times 10^{-4}$ & $1 \times 10^{-7}$ & \multicolumn {4} {c} {same as g, but absent  at $P<10$ mbar}  \\
[1ex]
\hline
\end{tabular}
\label{table1}
\end{table} 

\end{document}